\documentclass[nojss,shortnames]{jss}
\usepackage{amsmath}
\usepackage{amsfonts}
\usepackage{graphicx}
\usepackage{caption}
\usepackage{subcaption}
\usepackage{booktabs}
\usepackage{xspace}
\usepackage{subdepth}
\usepackage{float}

\usepackage{soul} 

\def\btheta{\mbox{\boldmath $\theta$}}
\def\bxi{\mbox{\boldmath $\xi$}}
\def\bkappa{\mbox{\boldmath $\kappa$}}
\def\bmu{\mbox{\boldmath $\mu$}}
\def\bSigma{\mathbf{\Sigma}}
\def\bnabla{\boldsymbol{\nabla}}
\def\by{\mathbf{y}}
\def\bs{\mathbf{s}}
\def\bA{\mathbf{A}}
\def\bB{\mathbf{B}}
\def\bI{\mathbf{I}}
\def\bS{\mathbf{S}}
\def\bX{\mathbf{X}}
\def\bD{\mathbf{D}}
\def\bR{\mathbf{R}}
\def\bx{\mathbf{x}}
\def\diag{\mbox{diag}}
\def\qmo{``}

\def\qmcsp{'' }

\newcommand{\suchthat}{\;\ifnum\currentgrouptype=16 \middle\fi|\;}

\providecommand\Student{\ensuremath{\text{Student--}t}\xspace}
\providecommand\SStudent{\ensuremath{\text{Skew--Student--}t}\xspace}

\providecommand{\V}{\mathsf{V}}
\providecommand{\bydef}{\ensuremath{\equiv}\xspace}

\author{%
David Ardia\\University of Neuch\^atel\\Laval University \And
Kris Boudt\\ Vrije Universiteit Brussel\\Vrije Universiteit Amsterdam\\\And
Leopoldo Catania\\University of Rome\\ \qmo Tor Vergata\qmcsp}
\Plainauthor{David Ardia, Kris Boudt, Leopoldo Catania} 

\title{Generalized Autoregressive Score Models in \proglang{R}:\\ The \pkg{GAS} Package}
\Plaintitle{Generalized Autoregressive Score Models in R: The GAS Package} 
\Shorttitle{The \proglang{R} package \pkg{GAS}} 

\Abstract{%
  This paper presents the \proglang{R} package \pkg{GAS} for the analysis of time series under the Generalized Autoregressive Score (GAS) framework of
  \citet{creal_etal.2013} and \citet{harvey.2013}. The distinctive feature of the GAS approach is the use of the score function as the driver of time--variation
   in the parameters of nonlinear models. The \pkg{GAS} package provides functions to simulate univariate and multivariate GAS processes, estimate the GAS parameters and to make time series forecasts. We illustrate the use of the \pkg{GAS} package with a detailed case study on estimating the time--varying conditional densities of a set of financial assets.
}
\Plainkeywords{GAS, time series models, score models, dynamic conditional score, R software}

\Address{
David Ardia \\
  Institute of Financial Analysis\\
  University of Neuch\^atel, Switzerland\\
  and\\
  Department of Finance, Insurance and Real Estate\\
  Laval University, Canada\\
  E-mail: \email{david.ardia@unine.ch}\\
  \\
  Kris Boudt\\
   Vrije Universiteit Brussel, Belgium\\
   and\\
  Vrije Universiteit Amsterdam, The Netherlands\\
  E-mail:  \email{kris.boudt@vub.ac.be}\\
  \\
  Leopoldo Catania (corresponding author)\\
  Department of Economics and Finance\\
  Faculty of Economics\\
  University of Rome, \qmo Tor Vergata\qmcsp\\
  Via Columbia, 2\\
  00133 Rome, Italy\\
  E-mail: \email{leopoldo.catania@uniroma2.it}\\
}

\begin{document}
\newpage

\section[introduction]{Introduction}\label{sec:intro}

Time--variation in the parameters describing a stochastic time series process is pervasive in almost all applied scientific fields. Early references to time series models include \citet{kalman.1960} and \citet{box_jenkins.1970}. In many settings, the model of interest is characterized by time--varying parameters, for which the literature has proposed a myriad of possible specifications. Recently, \citet{creal_etal.2013} and \citet{harvey.2013} note that many of the proposed models are either difficult to estimate (in particular, the class of stochastic volatility models reviewed in \cite{shephard.2005}) and/or do not properly take the shape of the conditional distribution of the data into account.\footnote{A typical example is the class of (G)ARCH models in which the squared (demeaned) return is the driver of time--variation in the conditional variance, independently of the shape of the conditional distribution of the return. To see that this is counter--intuitive, consider the case of observing a $10\%$ return when the conditional mean is $0\%$ and the volatility is $3\%$. Under the assumption of a normal distribution, the $10\%$ return is a strong signal of an increase in volatility, while under a fat--tailed \Student distribution, the signal is weakened because of the higher probability that the extreme value is an observation from the tails.} \citet{creal_etal.2013} and \citet{harvey.2013} therefore propose to use the score of the conditional density function as the main driver of time--variation in the parameters of the time series process used to describe the data. A further advantage of using the conditional score as driver  is that the estimation by Maximum Likelihood is straightforward. The resulting model is referred to as: Score--Driven model, Dynamic Conditional Score (DCS) model, or Generalized Autoregressive Score (GAS) model. In this article and accompanying \proglang{R} package, we use the GAS acronym for both.

The \proglang{R} package \pkg{GAS} is conceived to be of relevance for the modelling of all types of time series data. It does not matter whether they are real--valued, integer--valued, (0,1)--bounded or strictly positive, as long as there is a conditional density for which the score function and the Hessian are well--defined. The practical relevance of the GAS framework has been illustrated in the case of financial risk forecasting (see \emph{e.g.}, \citet{harvey_sucarrat.2014} for market risk, \citet{oh_patton.2013} for systematic risk, and \citet{creal_etal.2014} for credit risk analysis), dependence modelling (see \emph{e.g.}, \citet{harvey_thiele.2014} and \citet{janus_etal.2014}), and spatial econometrics (see \emph{e.g.}, \citet{blasques_etal.2014a} and \citet{catania_bille.2016}). For a more complete overview of the work on GAS models, we refer the reader to the GAS community page at \href{http://www.gasmodel.com/}{http://www.gasmodel.com/}.

It is important to note that, even though the GAS framework has been developed by econometricians, it is flexible enough to be used in all fields in which the use of time--varying parameter models is relevant. The main difficulty in using GAS models is to derive the score and Hessian and implementing the Maximum Likelihood estimation of the resulting nonlinear models. The \proglang{R} package \pkg{GAS} answers these needs by proposing an integrated set of \proglang{R} functions to do time series analysis in the \proglang{R} statistical language \citep{R} under the  GAS  framework. The functionalities include: (i) estimation, (ii) prediction, (iii) simulation, (iv) backtesting, and (v) graphical representation of the results, implying that it is ready to use in real--life applications. The user interface uses the \proglang{R} programming language, which has the advantage of being free and open source. However, most of the underlying routines are principally written in \proglang{C++} exploiting the \pkg{armadillo} library \citep{sanderson.2010} and the \proglang{R} packages \pkg{Rcpp} \citep{Rcpp.2011,Rcpp} and \pkg{RcppArmadillo} \citep{RcppArmadillo.2014,RcppArmadillo} to speed up the computations. Furthermore, since the package is written with the \proglang{S4} methods, \proglang{R} users with basic programming knowledge will find common functions such as \code{coef()}, \code{plot()} and \code{show()} to extract and analyze their results. We believe that this aspect is of primary importance since it dramatically increases the number of potential users. The \proglang{R} package \pkg{GAS} is available from the CRAN repository at \url{https://cran.r-project.org/package=GAS}. Other codes available for specific GAS models are available in the GAS community page at \href{http://www.gasmodel.com/}{http://www.gasmodel.com/}. For instance, the \proglang{R} package \pkg{betategarch} \citep{sucarrat.2013} allows us to estimate the beta--t--EGARCH model of \cite{harvey.2013} and its skewed version introduced by \cite{harvey_sucarrat.2014}.

The outline of the paper is as follows: Section~\ref{sec:models} reviews the GAS framework to define time--varying parameter	 models, referring to the seminal works of \citet{creal_etal.2013} and \citet{harvey.2013}. Section~\ref{sec:package} introduces the \proglang{R} package \pkg{GAS} and illustrates how to to simulate, estimate and make predictions. Section~\ref{sec:application} presents a real--life application to financial data. Section~\ref{sec:conclusion} concludes.

\section[The GAS framework to modeling time--varying parameters]{The GAS framework to modeling time--varying parameters}\label{sec:models}

One of the most appealing characteristics of the GAS framework is its applicability to define time--varying parameter models in a large variety of univariate and multivariate time series settings. We try to be as general as possible in reviewing the GAS framework, and report in Appendix~\ref{sec:gas_t} the detailed equations for the specific case of a conditionally \Student distributed random variable. In this section, we first introduce the notation and present the GAS model when the parameter space is unrestricted. We then show how a mapping function can be used to model the time--variation in the parameters when the parameter space is restricted. The section concludes by summarizing the Maximum Likelihood approach for GAS model estimation.

\subsection[Model specification]{Model specification}\label{sec:notation}

Let $\by_t\in\Re^N$ be an $N$--dimensional random vector at
time $t$ with conditional distribution:
\begin{equation}~\label{eq:dist}
  \by_t\vert\by_{1:t-1} \sim p(\by_t;\btheta_t) \,,
\end{equation}
where $\by_{1:t-1}\bydef(\by_1',\dots,\by_{t-1}')'$ contains the past values of $\by_t$ up to time $t-1$ and $\btheta_t\in\Theta\subseteq\Re^J$ is a vector of time--varying parameters which fully characterizes $p(\cdot)$ and only depends on $\by_{1:t-1}$ and a set of static additional parameters $\bxi$, \emph{i.e.}, $\btheta_t \bydef \btheta(\by_{1:t-1},\bxi)$ for all $t$. The main feature of GAS models is that the evolution in the time--varying parameter vector $\btheta_t$ is driven by the score of the conditional distribution defined in~\eqref{eq:dist}, together with an autoregressive component:
\begin{equation}~\label{eq:updating}
\btheta_{t+1} \bydef \bkappa + \bA \, \bs_t + \bB\,\btheta_t \,,
\end{equation}
where, $\bkappa$, $\bA$ and $\bB$ are matrices of coefficients with proper dimensions collected in $\bxi$, and $\bs_t$ is a vector which is proportional to the score of~\eqref{eq:dist}:
\begin{equation*}
\bs_t \bydef \bS_t(\btheta_t) \, \bnabla_t (\by_t,\btheta_t) \,.
\end{equation*}
The matrix $\bS_t$ is a $J\times J$ positive definite scaling matrix known at time $t$ and:
\begin{equation*}
  \bnabla_t(\by_t,\btheta_t) \bydef \frac{\partial\log p(\by_t;\btheta_t)}{\partial\btheta_t} \,,
\end{equation*}
is the score of~\eqref{eq:dist} evaluated at $\btheta_t$. \citet{creal_etal.2013} suggest to set the scaling matrix $\bS_t$  to a power $\gamma>0$ of the inverse of the Information Matrix of $\btheta_t$ to account for the variance of $\bnabla_t$. More precisely:
\begin{equation*}
  \bS_t(\btheta_t) \bydef \mathcal{I}_t(\btheta_t)^{-\gamma} \,,
\end{equation*}
with:
\begin{equation}~\label{eq:information_matrix}
  \mathcal{I}_t(\btheta_t) \bydef \E_{t-1}\left[\bnabla_t(\by_t,\btheta_t)\bnabla_t(\by_t,\btheta_t)'\right] \,,
\end{equation}
where the expectation is taken with respect to the conditional distribution of $\by_t$ given $\by_{1:t-1}$. The additional parameter $\gamma$ is fixed by the user and usually takes value in the set $\{0,\tfrac{1}{2},1\}$. When $\gamma=0$, $\bS_t=\bI$ and there is no scaling.\footnote{We denote by $\bI$ the identity matrix of appropriate size.} If $\gamma=1$ ($\gamma=\tfrac{1}{2}$), the conditional score $\bnabla_t(\by_t,\btheta_t)$ is premultiplied by the inverse of (the square root of) its covariance matrix $\mathcal{I}_t(\btheta_t)$.

It is worth noting that, whatever the choice of $\gamma$, $\bs_t$ is a Martingale Difference (MD) with respect to the distribution of $\by_t$ given $\by_{1:t-1}$, \emph{i.e.}, $\E_{t-1}\left[\bs_t\right]=\boldsymbol{0}$ for all $t$. Furthermore, when $\gamma = \tfrac{1}{2}$, the additional moment condition $\V_{t-1}\left[\bs_t\right]=\bI$ can be easily derived. Due to the fact that $\bs_t$ is a MD, if the spectral radius of $\bB$ is less then one\footnote{The spectral radius of a $L\times L$ matrix $\bX$ is defined as $\tau\left(\bX\right)\bydef\max\left(\vert\tau_1\vert, \dots, \vert\tau_L\vert\right)$, where $\tau_i$ is the $i$--th eigenvalue of $\bX$, for $i=1,\dots,L$.}, the updating equation of $\btheta_t$ reported in~\eqref{eq:updating} implies a mean reverting process for $\btheta_t$ through the long--term mean $\left(\bI - \bB\right)^{-1}\bkappa$, which means that the unconditional value of $\btheta_t$ is $\left(\bI - \bB\right)^{-1}\bkappa$. It follows that, the $J$--valued vector $\bkappa$ and the $J\times J$ matrix $\bB$ control for the level and the persistence of the process, respectively.

The additional $J\times J$ matrix of coefficients $\bA$, that premultiplies the scaled score $\bs_t$, controls for the impact of $\bs_t$ to $\btheta_{t+1}$. Specifically, as detailed in \cite{creal_etal.2013}, the quantity $\bs_t$ indicates the direction to update the vector of parameters from $\btheta_t$, to $\btheta_{t+1}$, acting as a steepest ascent algorithm for improving the model local fit given the current parameter position. Interestingly, this updating procedure resembles the well--known Newton--Raphson algorithm. Hence, $\bA$ can be interpreted as the step of the update, and needs to be designed in a way to do not distort the signal coming from $\bs_t$; see Section~\ref{sec:ml_gas}.

\subsection[Reparametrization]{Reparametrization}\label{sec:reparametrisation}

In~\eqref{eq:updating} the parameter vector $\btheta_t$ has a linear specification and is thus unbounded. In practice, the parameter space of $\btheta_t$ is often restricted ($\Theta\subset\Re^J$).  For instance, when we model the scale parameter of a \Student distribution, we need to ensure its positiveness. Even if this problem can be solved by imposing constraints on $\bxi$ \cite[as is done in the GARCH model, see,][]{bollerslev.1986}, the standard solution under the GAS framework is to use a (possibly nonlinear) link function $\Lambda(\cdot)$ that maps $\widetilde\btheta_t\in\Re^J$ into $\btheta_t$ and where $\widetilde\btheta_t\in\Re^J$ has the linear dynamic specification of~\eqref{eq:updating}.
Specifically, let $\Lambda:\Re^J\to\Theta$ be a twice differentiable vector--valued mapping function such that $\Lambda(\widetilde\btheta_t) = \btheta_t$. The updating equation for $\btheta_t$ is then given by:
\begin{align}\label{eq:updating_rep}
\begin{split}
  \btheta_t &\bydef \Lambda(\widetilde\btheta_t)\\
  \widetilde\btheta_t &\bydef \bkappa + \bA\widetilde\bs_t + \bB\widetilde\btheta_{t-1} \,,
\end{split}
\end{align}
where $\widetilde\bs_t \bydef \widetilde \bS_t(\widetilde\btheta_t) \,\widetilde\bnabla_t(\by_t,\widetilde\btheta_t)$ and $\widetilde\bnabla_t(\by_t,\widetilde\btheta_t)$ represents the score of~\eqref{eq:dist} with respect to $\widetilde\btheta_t$, and, consequently, $\widetilde \bS_t(\widetilde\btheta_t)$ can depend on the information matrix of $\widetilde\btheta_t$ given by $\mathcal{\widetilde I}_t(\widetilde\btheta_t)$. Denote the Jacobian matrix of $\Lambda(\cdot)$ evaluated at $\widetilde\btheta_t$ as follows:
\begin{equation*}
  \mathcal{J}(\widetilde\btheta_t) \bydef \left.\frac{\partial\Lambda(\widetilde\btheta_t)}{\partial\widetilde\btheta_t}\right.\,.
\end{equation*}
Then, the following relations hold:
\begin{align*}
  \widetilde\bnabla_t(\by_t,\widetilde\btheta_t) &= \mathcal{J}(\widetilde\btheta_t)'\bnabla_t(\by_t,\btheta_t) \\
  \mathcal{\widetilde I}_t(\widetilde\btheta_t) &= \mathcal{J}(\widetilde\btheta_t)'\mathcal{I}_t(\btheta_t)\mathcal{J}(\widetilde\btheta_t) \,.
\end{align*}
 This way, almost all the nonlinear constraints can be easily handled via the definition of a proper mapping function $\Lambda(\cdot)$ and its associated Jacobian matrix $\mathcal{J}(\cdot)$. The coefficients to be estimated are gathered into $\bxi \bydef (\bkappa, \bA, \bB)$ and estimated by numerically maximizing the (log-)likelihood function as detailed in Section~\ref{sec:ml_gas}.\footnote{Clearly, the coefficients $\bkappa$, $\bA$ and $\bB$ in~\eqref{eq:updating_rep} are different from those of~\eqref{eq:updating}, however, for notational purposes, we continue to use the same notation.} In Appendix~\ref{sec:mapping} we discuss the choice of mapping function for GAS models in more details.

\subsection[Maximum likelihood estimation]{Maximum likelihood estimation}\label{sec:ml_gas}

A useful property of GAS models is that, given the past information and the static parameter vector $\bxi$, the vector of time--varying parameters, $\btheta_t$, is perfectly predictable and the loglikelihood function can be easily evaluated via the prediction error decomposition. More precisely, for a sample of $T$ realizations of $\by_t$, collected in $\by_{1:T}$, the vector of parameters $\bxi$ can be estimated by Maximum Likelihood (ML) as the solution of:
\begin{equation}\label{eq:llk_max}
  \widehat{\bxi} \bydef \underset{\bxi}{\arg\max}~\mathcal{L}\left(\bxi;\by_{1:T}\right),
\end{equation}
where:
\begin{equation*}
  \mathcal{L}\left(\bxi;\by_{1:T}\right) \bydef \log p\left(\by_{1};\btheta_1\right) + \sum_{t=2}^{T}\log p\left(\by_{t};\btheta_t\right)\,,
\end{equation*}
 $\btheta_1 \bydef (\bI - \bB)^{-1}\bkappa$, and, for $t>1$, $\btheta_t \bydef \btheta(\by_{1:t-1},\bxi)$. Note the dependence of $\btheta_t$ on $\bxi$ and $\by_{1:t-1}$.

There are two important caveats in the ML evaluation of GAS models. The first one is that, from a theoretical perspective, ML estimation of GAS models is an on--going research topic. General results are reported by \citet{harvey.2013}, \cite{blasques_etal.2014b} and \citet{blasques_etal.2014c}, while results for specific models have been derived by \citet{andres.2014} and \citet{blasques_etal.2014a}.

The second one is that, even when the ML estimator is consistent and asymptotically normal, the numerical maximization of the loglikelihood function in~\eqref{eq:llk_max} can be challenging, because of the nonlinearities induced by $\Lambda\left(\cdot\right)$ and the way $\by_t$ enters the scaled score $\bs_t$. Consequently, when the optimizer is gradient--based, good starting values need to be selected for GAS models. In the \proglang{R} package \pkg{GAS}, starting values for the optimizer are chosen in the following way: (i) estimate the static version of the model (\emph{i.e.}, with $\bA=\boldsymbol{0}$ and $\bB=\boldsymbol{0}$) and set the initial value of $\bkappa$ accordingly, and (ii) perform a grid search for the coefficients contained in $\bA$ and $\bB$. Further technical details are presented in Section~\ref{sec:gas_est}.

Implementation of the models in the \proglang{R} package \pkg{GAS} follows the common approach in the GAS literature. First, matrices $\bA$ and $\bB$ are constrained to be diagonal. Second, in order to avoid an explosive pattern for $\widetilde\btheta_t$, the spectral radius of $\bB$ is constrained to be less than one. Third, the positiveness of each element of $\bA$ is imposed in order to do not distort the signal coming from the conditional score $\bs_t$.

\section[The R package GAS]{The \proglang{R} package \pkg{GAS}}\label{sec:package}

The \proglang{R} package \pkg{GAS} offers an integrated environment to deal with GAS models in \proglang{R}. Its structure is somehow similar to the \proglang{R} package \pkg{rugarch} \citep{rugarch} for GARCH models, which is widely used by practitioners and academics. The similarities concern the steps the user has to do to perform her analysis as well as the type of functions she faces. Specifically, the first step is to specify the model, which means choosing: (i) the assumptions for the conditional distribution of the data, (ii) the set of parameters that have to vary over time and, (iii) the scaling mechanism for the conditional score. These steps are detailed in Section~\ref{sec:gas_spec}. Once the model is properly specified, the user can estimate the unknown parameters in $\bxi$ by numerical maximization of the log--likelihood function as detailed in Section~\ref{sec:gas_est}. Finally, predictions according to the estimated model can be easily performed; see Section~\ref{sec:gas_fore}. Simulation of GAS models is presented in Section~\ref{sec:gas_sim}.

Functions for: (i) specification, (ii) estimation, (iii) forecasting and (iv) simulation are available for univariate and multivariate time series. The general nomenclature for the functions when we consider univariate time series is \qmo\code{UniGAS\dots()}\qmcsp and that for multivariate time series is \qmo\code{MultiGAS\dots()}\qmcsp.

In the \proglang{R} package \pkg{GAS}, several datasets are also included for reproducibility
purposes, such as: US inflation (\code{cpichg}), US unemployment rate (\code{usunp}), realized volatility of the S\&P500 Index (\code{sp500rv}) and intraday bid and ask quotes for Citygroup corporation (\code{tqdata}). These datasets are freely available online; see the \proglang{R} documentation for references. In this section, we use the monthly US inflation measured as the logarithmic change in the CPI available from the Federal Reserve Bank of St. Louis website \href{https://fred.stlouisfed.org/}{https://fred.stlouisfed.org/}. This dataset can be easily loaded in the \proglang{R} workspace using: \code{data("cpichg", package = "GAS")}.

\subsection[Specification]{Specification}\label{sec:gas_spec}

Specification of GAS models is the first step the user needs to undertake. This is achieved by using the \code{UniGASSpec()} and \code{MultiGASSpec()} functions, in the cases of univariate and multivariate models, respectively. Both functions accept three arguments and return an object of the class \code{uGASSpec} and \code{mGASSpec}, respectively. The three arguments are:
\begin{itemize}
  \item[-] \code{Dist}: A \code{character} indicating the label of the conditional distribution assumed for the data. Available distributions can be displayed using the function \code{DistInfo()} and are reported in Table~\ref{tab:dist}. By default \code{Dist = "norm"}, \emph{i.e.}, the Gaussian distribution.
  \item[-] \code{ScalingType}: A \code{character} indicating the scaling mechanism for the conditional score, \emph{i.e.}, the value of the $\gamma$ parameter in~\eqref{eq:information_matrix}. Possible choices are \code{"Identity"} ($\gamma = 0$), \code{"Inv"} ($\gamma = 1$) and \code{"InvSqrt"} ($\gamma = \tfrac{1}{2}$). Note that, for some distributions only \code{ScalingType = "Identity"} is supported; see function \code{DistInfo()} and Table~\ref{tab:dist}. By default \code{ScalingType = "Identity"}, \emph{i.e.}, no scaling occurs.\footnote{In the \proglang{R} package \pkg{GAS} the information matrices and the scores are always computed using their analytical formulations.}
  \item[-] \code{GASPar}: A named \code{list} with \code{boolean} entries containing information about which parameters of the conditional distribution have to be time--varying. Generally, each univariate distribution is identified by a series of maximum four parameters. These are indicated by \code{location, scale, skewness} and \code{shape}. Note that, for some distributions, these labels are not strictly related to their literal statistical meaning. Indeed, for the Exponential distribution \code{exp}, the term \code{location} indicates the usual intensity rate parameter; see the \code{DistInfo()} function for more details. For multivariate distributions, the set of parameters is indicated by \code{location, scale, correlation} and \code{shape}\footnote{In the \proglang{R} documentation an extra parameter, \code{shape2}, is reported. This will be used for future extensions of the package.}. For example, in the case of a multivariate \Student distribution with mean vector $\bmu_t$, scale matrix $\bSigma_t \bydef \bD_t\bR_t\bD_t$, where $\bD_t$ is the diagonal matrix of scales and $\bR_t$ is the correlation matrix, and $\nu_t$ degrees of freedom, we have that: \code{location} refers to $\bmu_t$, \code{scale} refers to $\bD_t$, \code{correlation} refers to $\bR_t$ and \code{shape} refers to $\nu_t$. By default, \code{GASPar = list(location = FALSE, scale = TRUE, skewness = FALSE, shape = FALSE)} for the univariate case, and \code{GASPar = list(location = FALSE, scale = TRUE, correlation = FALSE, shape = FALSE)} for the multivariate case.
\end{itemize}

\begin{table}[!t]
	\centering
	\resizebox{1.0\columnwidth}{!}{%
		\begin{tabular}{ccccccc}
			\toprule
			Label & Name & Type & Parameters & \# & Scaling Type \\
			\hline
            \code{norm} & Gaussian & univariate & \code{location, scale} & 2 & \code{Identity, Inv, InvSqrt} \\
\code{std} & \Student (i) & univariate & \code{location, scale, shape} & 3 & \code{Identity, Inv, InvSqrt} \\
\code{sstd} & \SStudent (ii) & univariate & \code{location, scale, skewness, shape} & 4 & \code{Identity} \\
\code{ald} & Asymmetric Laplace (iii) & univariate & \code{location, scale, skewness} & 3 & \code{Identity, Inv, InvSqrt} \\
\code{poi} & Poisson (iv)& univariate & \code{location} & 1 & \code{Identity, Inv, InvSqrt} \\
\code{gamma} & Gamma & univariate & \code{scale, shape} & 2 & \code{Identity, Inv, InvSqrt} \\
\code{exp} & Exponential (v)& univariate & \code{location} & 1 & \code{Identity, Inv, InvSqrt} \\
\code{beta} & Beta (vi)& univariate & \code{scale, shape} & 2 & \code{Identity, Inv, InvSqrt} \\
\code{mvnorm} & Multivariate Gaussian & multivariate & \code{location, scale, correlation} & 9 (vii) & \code{Identity} \\
\code{mvt} & Multivariate \Student & multivariate & \code{location, scale, correlation, shape} & 10 (vii) & \code{Identity} \\
			\bottomrule
		\end{tabular}
	}
	\caption{Statistical distributions or which the  \proglang{R} package \pkg{GAS} provides the functionality to simulate, estimate and forecast the time--variation in its parameters. The fifth column, \#, reports the number of parameters of the distribution. Note: (i) the usual \Student distribution (not reparametrised in terms of the variance parameter),
(ii)  the reparametrised \SStudent such that the location and scale parameters coincide with the mean and the standard deviation of the distribution as done in the \pkg{rugarch} package, (iii) the \code{ald} distribution used the $\theta$, $\sigma$ and $\kappa$ reparametrization, as specified in \citet{kotz_etal.2001}, (iv) for the Poisson distribution \code{location} means the usual intensity parameter, (v) for the Beta distribution \code{shape} means the usual $\alpha$ parameter and \code{scale} means the usual $\beta$ parameter, (vi) for the Exponential distribution \code{location} means the usual rate parameter, (vii) for $N=3$.}
\label{tab:dist}
\end{table}

The function \code{MultiGASSpec()} also accepts the additional boolean argument \code{ScalarParameters} controlling for the parametrization of $\bA$ and $\bB$ in~\eqref{eq:updating_rep}. Setting \code{ScalarParameters = TRUE} (the default value), the coefficients controlling the evolution of the location, scale and correlation parameters are constrained to be the same across each group. Specifically, if $\by_t\in\Re^3$ follows a GAS process with conditional multivariate Gaussian distribution, the vector of time--varying parameters is $\btheta_t = \left(\mu_{1,t}, \mu_{2,t}, \mu_{3,t}, \sigma_{1,t}, \sigma_{2,t}, \sigma_{3,t}, \rho_{21,t}, \rho_{31,t},\rho_{32,t}\right)^\prime$. If \code{ScalarParameters = TRUE}, the matrix of coefficients $\bA$ is parameterized as:
\begin{equation*}
\bA \bydef \diag\left(a_\mu, a_\mu, a_\mu, a_\sigma, a_\sigma, a_\sigma, a_\rho, a_\rho, a_\rho\right) \,,
\end{equation*}
while, if \code{ScalarParameters = FALSE}, the matrix of coefficients $\bA$ takes the form:
\begin{equation*}
\bA \bydef \diag\left(a_{\mu_1}, a_{\mu_2}, a_{\mu_3}, a_{\sigma_1}, a_{\sigma_2}, a_{\sigma_3}, a_{\rho_{21}}, a_{\rho_{31}}, a_{\rho_{32}}\right) \,.
\end{equation*}
Hence, in the latter case, each element of $\btheta_t$ evolves heterogeneously with respect to the others. The same constraints are applied to $\bB$, which means that, if \code{ScalarParameters = TRUE}, for the general $N$ case, the number of parameters decreases from $3N\left(N+1\right)/2$ to $N\left(N+1\right)/2 + 2$. Additional constraints are introduced through the \code{GASPar} argument as in the univariate case; see \code{help("MultiGASSpec")}.

As an illustration, assume that we want to specify a \Student GAS model with time--varying conditional mean and scale parameters, but fixed degree of freedom, \emph{i.e.}, $\nu_t = \nu$. This can be easily done with the following lines of code:
\begin{CodeChunk}
\begin{CodeInput}
R> GASSpec <- UniGASSpec(Dist = "std", ScalingType = "Identity",
                        GASPar = list(location = TRUE, scale = TRUE,
                                      shape = FALSE))
\end{CodeInput}
\end{CodeChunk}
Details about the object returned from \code{UniGASSpec()} are printed in the console by simply calling \code{GASSpec}:
\begin{CodeChunk}
\begin{CodeInput}
R> GASSpec
\end{CodeInput}
\begin{CodeOutput}
-------------------------------------------------------
-            Univariate GAS Specification             -
-------------------------------------------------------
Conditional distribution
-------------------------------------------------------
Name: Student-t
Label: std
Type: univariate
Parameters: location, scale, shape
Number of Parameters: 3
References:	
-------------------------------------------------------
GAS specification
-------------------------------------------------------
Score scaling type:  Identity
Time varying parameters:  location, scale
-------------------------------------------------------
\end{CodeOutput}
\end{CodeChunk}
Since the scaling matrix $\bS_t$ is set to the identify matrix  (\emph{i.e.}, \code{ScalingType = "Identity"}) this model for the conditional \Student distribution corresponds to the one described in Appendix~\ref{sec:gas_t}. Multivariate GAS specifications are analogously specified using the \code{MultiGASSpec()} function; see \code{help("MultiGASSpec")}.

\subsection[Estimation]{Estimation}\label{sec:gas_est}

Similar to model specification, estimation is handled with two different functions for univariate and multivariate models: \code{UniGASFit()} and \code{MultiGASFit()}, respectively. These functions require only two arguments: the GAS specification object \code{GASSpec} and the data, and returns an object of the class \code{uGASFit} or \code{mGASFit}. As an example, let us estimate the GAS model previously specified using the US inflation data included in the \proglang{R} package \pkg{GAS}:
\begin{CodeChunk}
\begin{CodeInput}
R> data("cpichg", package = "GAS")
R> Fit <- UniGASFit(GASSpec, cpichg)
\end{CodeInput}
\end{CodeChunk}
The computational time is less than one second on a modern computer. The optimizer used is the General Nonlinear Augmented Lagrange Multiplier method of \cite{ye.1988} available in the \proglang{R} package \pkg{Rsolnp} \citep{Rsolnp}. Results can be inspected by calling the object \code{Fit}.
\begin{CodeChunk}
\begin{CodeInput}
R> Fit
\end{CodeInput}
\begin{CodeOutput}
------------------------------------------
-          Univariate GAS Fit            -
------------------------------------------

Model Specification:	
T =  276
Conditional distribution:  std
Score scaling type:  Identity
Time varying parameters:  location, scale
------------------------------------------
Estimates:
       Estimate Std. Error t value  Pr(>|t|)
kappa1  0.03735    0.02991   1.249 1.059e-01
kappa2 -0.25994    0.13553  -1.918 2.755e-02
kappa3  0.92671    0.76127   1.217 1.117e-01
a1      0.07173    0.01780   4.030 2.787e-05
a2      0.45377    0.20828   2.179 1.468e-02
b1      0.94318    0.02600  36.272 0.000e+00
b2      0.85559    0.07185  11.908 0.000e+00

------------------------------------------
Unconditional Parameters:
location    scale    shape
  0.6574   0.1653   6.5262

------------------------------------------
Information Criteria:
   AIC    BIC     np    llk
 370.4  395.8    7.0 -178.2
 ------------------------------------------

Elapsed time: 0.01 mins
\end{CodeOutput}
\end{CodeChunk}
The output printed in the console is divided into: (i) the summary of the model, (ii) the estimated coefficients along with significance levels according to their asymptotic normal distribution, (iii) the unconditional level of the parameters, \emph{i.e.}, $\Lambda((\bI - \widehat{\bB})^{-1}\widehat{\bkappa})$, (iv) AIC and BIC information criteria in addition to the number of estimated parameters (\code{np}), the log--likelihood (\code{llk}) evaluated at its optimum, and (v) the computation time.

Concerning the estimated coefficients, \code{kappa1}, \code{kappa2} and \code{kappa3} are the elements of vector $\bkappa$ in~\eqref{eq:update_studentt}, \emph{i.e.}, $\kappa_\mu, \kappa_\phi$ and $\kappa_\nu$, respectively. Analogously, \code{a1} and \code{a2} are the estimates of $a_\mu$ and $a_\phi$ and \code{b1} and \code{b2} are estimates of $b_\mu$ and $b_\phi$, where $\phi$ refers to the scale parameter of the \Student distribution; see Appendix~\ref{sec:gas_t}. Note that, since we have specified \code{scale = FALSE} in the \code{UniGASSpec()} function, coefficients \code{a3} and \code{b3}, corresponding to $a_\nu$ and $b_\nu$ are not reported (and constrained to zero during the optimization).

The \proglang{R} package \pkg{GAS} provides several methods to extract the relevant estimated quantities for objects of the class \code{uGASFit} or \code{mGASFit}. They allow us to: (i) calculate several quantities of the estimated conditional distribution at each point in time, such as: quantiles, conditional moments and filtered parameters (see \code{quantile(Fit)}, \code{getMoments(Fit)} and \code{getFilteredParameters(Fit)}, respectively), (ii) extract the estimated coefficients (\code{coef(Fit)}), (iii) generate a graphical representation of the results (\code{plot(Fit)}); see \code{help("uGASFit")} for details.

\subsection[Forecasting]{Forecasting}\label{sec:gas_fore}

Forecasting is a crucial aspect in applied time series analysis. Given the parametric assumption of GAS models, predictions are usually given in the form of density forecasts, \emph{i.e.}, the distribution of $\by_{T+h}\vert\by_{1:T}$ for $h\geq1$. Knowing the predictive density, practitioners can extract any relevant quantities such as future expected value $\E_T\left[\by_{T+h}\right]$ or (co--)variance  $\V_T\left[\by_{T+h}\right]$. For GAS models, the one--step ahead predictive distribution ($h=1$) is analytically available while it needs to be estimated
by simulation in the multi--step ahead case ($h>1$).

The \proglang{R} package \pkg{GAS} can handle both one--step and multi--step ahead forecasts. Consistent with previous nomenclature, functions for univariate and multivariate predictions are \code{UniGASFor()} and \code{MultiGASFor()}, respectively. These
functions accept an object of the class \code{uGASFit} or \code{mGASFit}, created using the functions \code{UniGASFit()} and \code{MultiGASFit()}, and return an object of the class \code{uGASFor} and \code{mGASFor}, respectively. Additional arguments are:
\begin{itemize}
  \item \code{H}: a \code{numeric} integer value representing the forecast horizon, \emph{i.e.}, $h$. By default \code{H = 1}.
  \item \code{B}: a \code{numeric} integer value representing the number of draws to approximate the multi--step ahead predictive distribution when $h>1$. By default \code{B = 1e4}.
  \item \code{ReturnDraws}: a \code{boolean} controlling if the simulated draws from $\by_{T+1}\vert \by_{1:T}$, $\by_{T+2}\vert \by_{1:T},\dots$, $\by_{T+h}\vert \by_{1:T}$ have to be returned. By default \code{ReturnDraws = FALSE}.
\end{itemize}
Other arguments to perform rolling--type of forecasts are detailed in the documentation; see \code{help("UniGASFor")}. Practically, if we want to predict the next--year inflation (\emph{i.e.}, $h=12$ with the monthly series \code{cpichg}), after having estimated the GAS model of Section~\ref{sec:gas_est}, we can execute the following code:
\begin{CodeChunk}
\begin{CodeInput}
R> Forecast <- UniGASFor(Fit, H = 12)
\end{CodeInput}
\end{CodeChunk}
and inspect the results by calling the object \code{Forecast}:
\begin{CodeChunk}
\begin{CodeInput}
R> Forecast
\end{CodeInput}
\begin{CodeOutput}
------------------------------------------
-        Univariate GAS Forecast         -
------------------------------------------

Model Specification
Conditional distribution:  std
Score scaling type:  Identity
Horizon:  12
Rolling forecast:  FALSE
------------------------------------------
Parameters forecast:
    location  scale shape
T+1  0.10128 0.1524 6.526
T+2  0.09497 0.1737 6.526
T+3  0.09380 0.2151 6.526
T+4  0.09253 0.2577 6.526
T+5  0.08745 0.3020 6.526

....................
     location  scale shape
T+8   0.08345 0.4219 6.526
T+9   0.07790 0.4575 6.526
T+10  0.07380 0.4900 6.526
T+11  0.07557 0.5199 6.526
T+12  0.07507 0.5465 6.526
\end{CodeOutput}
\end{CodeChunk}
which returns some model information and the predictions of future model parameters based on averages over \code{B} draws. \code{Forecast} is an object of the class \code{uGASFor} and comes with several methods to extract and visualize the results; see \code{help("uGASFor")}.

As commonly done in time series analysis, predictions are generated from models fitted to rolling windows. The \proglang{R} package \pkg{GAS} includes this functionality via \code{UniGASRoll()} and \code{MultiGASRoll()}. These functions accept several arguments that we briefly describe in the univariate case:
\begin{itemize}
  \item \code{data}: a vector of length $T + $\code{ForecastLength} containing all the observations.
  \item \code{GASSpec}: an object of the class \code{uGASSpec} created with \code{UniGASSpec()}.
  \item \code{ForecastLength}: a \code{numeric} integer which specifies the length of the out--of--sample.
  \item \code{RefitEvery}: a \code{numeric} integer of periods before coefficients are re--estimated.
  \item \code{RefitWindow}: a \code{character} for the type of the window. As in the \proglang{R} package \pkg{rugarch}, we define the options: \code{RefitWindow = "recursive"} and \code{RefitWindow = "moving"}. If \code{RefitWindow = "recursive"} all past observations are used when the model is re--estimated. If \code{RefitWindow = "moving"} initial observations are eliminated.
\end{itemize}
Other arguments useful to tailor the forecasting procedure and to parallelize the code execution are available and detailed in the \proglang{R} documentation; see \code{help("UniGASRoll")}.

Suppose now we are interested in assessing the forecast performance of the GAS model with a \Student conditional distribution and time--varying location and scale parameters, detailed in Appendix~\ref{sec:gas_t}, and specified in the object \code{GASSpec} in Section~\ref{sec:gas_spec}. We treat the last 150 observations of \code{cpichg} as out--of--sample and run a rolling--window forecast exercise using the following portion of code:
\begin{CodeChunk}
\begin{CodeInput}
Roll <- UniGASRoll(cpichg, GASSpec, ForecastLength = 150,
                  RefitEvery = 3, RefitWindow = "moving")
\end{CodeInput}
\end{CodeChunk}
where model coefficients are re--estimated quarterly (\emph{i.e.}, every three observations with monthly data) using a moving windows (\code{RefitWindow = "moving"}). The code automatically makes a series of one--step ahead rolling predictions according to the model estimated using only the past information. This way, the user can perform out--of--sample analysis with GAS models. The object \code{Roll} belongs to the class \code{uGASRoll} which, as \code{uGASFit} and \code{uGASFor}, comes with several methods to extract and represent
the results; see \code{help("uGASRoll")}.

\subsection[Simulation]{Simulation}\label{sec:gas_sim}

Simulation of univariate and multivariate GAS models is straightforward with the \proglang{R} package \pkg{GAS}. This can be easily done via \code{UniGASSim()} and \code{MultiGASSim()}; see the \proglang{R} documentation. Several examples, also investigating the finite sample properties of the ML estimator for GAS models, are reported in the \code{inst/test/Simulation.R} file included in the package tarball.

Besides selecting the conditional distribution of the time series process, the user of course needs to specify also the static parameters $\bxi$ governing the dynamics in $\btheta_t$. More precisely, for simulation of GAS models, the vector $\bkappa$ and the matrices $\bA$ and $\bB$, need to be specified. It is worth stressing that, the definition of $\bkappa$ can be tricky, especially for multivariate models. The difficulty emerges from the fact that, $\bkappa$ determines the unconditional value of the reparametrised vector of parameters $\widetilde\btheta_t$, implying that, if the user wants to specify the model in terms of a target value $\btheta^*$ she needs to know the inverse of the mapping function $\Lambda\left(\cdot\right)$.\footnote{Here we define the \qmo target value\qmcsp as the unconditional level of the parameters the user has in mind. This targeting approach requires the time--varying parameter model to be stationary, as explained in, \emph{e.g.}, \cite{blasques_etal.2014e}.} To address this problem, the functions \code{UniUnmapParameters()} and \code{MultiUnmapParameters()}, representing $\Lambda^{-1}\left(\cdot\right)$, are available for univariate and multivariate models, respectively. This way, the user can easily specify $\bkappa$ such that $\bkappa^* \bydef (\bI - \bB)\Lambda^{-1}(\btheta^*)$. Table~\ref{tab:bounds} lists the numerical bounds imposed for the univariate distributions, such that \code{UniUnmapParameters()} cannot takes values outside those ranges. For the multivariate case, please refer to the examples reported in the \code{inst/test/SimulateGAS.R} file included in the package tarball.

\begin{table}[H]
	\centering
	\setlength\tabcolsep{25pt}
	\begin{tabular}{ccccc}
		\toprule
		Label & \code{location} & \code{scale} & \code{skewness} & \code{shape} \\
		\hline
		\code{norm} & $\Re^{~~}$ & $\Re^+$ & -- & --  \\
		\code{std} & $\Re^{~~}$ & $\Re^+$ & -- & $\left(2.01,50.0\right)$  \\
		\code{snorm} & $\Re^{~~}$ & $\Re^+$ & $\left(0.1,2.0\right)$ & --  \\
		\code{sstd} & $\Re^{~~}$ & $\Re^+$ & $\left(0.1,2.0\right)$ & $\left(2.01,50.0\right)$  \\
		\code{ald} & $\Re^{~~}$ & $\Re^+$ & $\Re^+$ & --  \\
		\code{poi} & $\Re^+$ & -- & -- & --  \\
		\code{gamma} & $\Re^+$ & $\Re^+$ & -- & --  \\
		\code{exp} & $\Re^+$ & -- & -- & --  \\
		\code{beta} & $\Re^+$ & $\Re^+$ & -- & --  \\
		\bottomrule
	\end{tabular}
	\caption{Overview of the restrictions on the allowed values for the parameters of the univariate distributions, for which the \proglang{R} package \pkg{GAS} provides the functionality to simulate, estimate and forecast the time variation in the parameters. When the parameter space is $\Re^+$, we use the exponential link function reported in~\eqref{eq:exponential_link} with $c=0$, while when the space is of the type $(a, b)$, we use the modified logistic link function
		reported in~\eqref{eq:logistic_link}; see Appendix~\ref{sec:mapping}.}
	\label{tab:bounds}
\end{table}

Suppose we want to simulate $T = 1,000$ observations from the \Student GAS model reported in Appendix~\ref{sec:gas_t} with time--varying location and scale, but constant shape parameters. Assume our target value for the parameters is $\btheta^* = \left(\mu^*, \sigma^*, \nu^*\right)^\prime$ with $\mu^* = 0.1, \sigma^* = 1.5$ and $\nu^* = 7$. The matrix $\bA$ and $\bB$ are defined as:
\begin{align*}
  \bA &= \diag\left(0.1, 0.4, 0.0\right)\\
  \bB &= \diag\left(0.9, 0.95, 0.0\right),
\end{align*}
such that both the conditional mean and the conditional variance evolve quite persistently over time, while the shape parameter is constant. The implementation of \code{UniUnmapParameters()} and \code{UniGASSim()} proceeds as:
\begin{CodeChunk}
\begin{CodeInput}
A <- diag(c(0.1, 0.4, 0.0))
B <- diag(c(0.9, 0.95, 0.0))
ThetaStar <- c(0.1, 1.5, 7.0)
Kappa <- (diag(3) - B) %*% UniUnmapParameters(ThetaStar, "std")

Sim <- UniGASSim(T = 1000, Kappa, A, B,
                 Dist = "std", ScalingType = "Identity")
\end{CodeInput}
\end{CodeChunk}
where \code{Sim} is an object of the class \code{uGASSim} and comes with several method such as \code{show}, \code{plot}, and \code{getMoments}, among others; see \code{help("uGASSim")}.

\section[Applications to financial data]{Applications to financial data}\label{sec:application}

In order to illustrate how the \proglang{R} package \pkg{GAS} can be used in practical situations, we present an empirical application with univariate and multivariate time series of financial returns. We consider daily log--returns (in percentage points) of the Dow Jones 30 constituents available in the \code{dji30ret} dataset. This dataset includes the closing value log--returns from March 3rd, 1987 to February 3rd, 2009 for a total of 5,521 observations per series. The dataset can be easily loaded in the workspace using:
\begin{CodeChunk}
\begin{CodeInput}
R> library("GAS")
R> data("dji30ret", package = "GAS")
\end{CodeInput}
\end{CodeChunk}
where \code{dji30ret} is a $5521 \times 30$ \code{data.frame} containing the daily log--returns. Our analysis is a typical out--of--sample exercise, meaning that: (i) we estimate the models using an in--sample period, (ii) we do predictions of the conditional distribution for the observations in the out--of--sample period, (iii) and that we compare the models according to their out--of--sample performance.

The models we consider are univariate/multivariate GAS models estimated with the \proglang{R} package \pkg{GAS}, and univariate/multivariate GARCH models estimated using the popular \proglang{R} packages \pkg{rugarch} \citep{rugarch} and \pkg{rmgarch}  \citep{rmgarch}, respectively. The univariate specifications we consider are: (i) the \SStudent GAS model with only time--varying scale parameter (\emph{i.e.}, \code{Dist = "sstd"}) and, (ii) the GARCH(1,1) model with \SStudent distributed error. For both models we employ the \SStudent distribution of \cite{fernandez_steel.1998} reparametrised such that the location and scale parameters coincide with the mean and the standard deviation of the distribution as done in the \pkg{rugarch} package.

For the multivariate specifications, we consider: (i) the GAS model with conditional multivariate \Student distribution with time--varying scales and correlations used in \citet{creal_etal.2011} and, (ii) the Dynamic Conditional Correlation (DCC) model of \citet{engle.2002} with a conditional multivariate \Student distribution. For simplicity, the multivariate analysis only considers three series of the whole dataset: Caterpillar Inc. (CAT), 3M (MMM) and Pfizer Inc. (PFE).

The code used to specify the univariate and multivariate GAS models is:
\begin{CodeChunk}
\begin{CodeInput}
R> uGASSpec <- UniGASSpec(Dist = "sstd",
                         ScalingType = "Identity",
                         GASPar = list(scale = TRUE))
\end{CodeInput}
\end{CodeChunk}
and:
\begin{CodeChunk}
\begin{CodeInput}
R> mGASSpec <- MultiGASSpec(Dist = "mvt",
                           ScalingType = "Identity",
                           GASPar = list(scale = TRUE,
                                         correlation = TRUE))

\end{CodeInput}
\end{CodeChunk}
respectively.

The last $H = 3,000$ observations (from January 27th, 1991, to the end of the sample) compose the out--of--sample period. During the out--of--sample period,
one--step ahead density predictions are constructed by the univariate and multivariate models. Models (and therefore coefficients) are re--estimated using a moving--window every hundredth observations, as detailed in Section~\ref{sec:gas_fore}. One--step ahead rolling prediction are then computed as:
\begin{CodeChunk}
\begin{CodeInput}
luGASRoll <- list()

N <- ncol(dji30ret)

for(i in 1:N){
   luGASRoll[[i]]  <- UniGASRoll(data = dji30ret[, i],
                                 GASSpec = uGASSpec,
                                 ForecastLength = 3000,
                                 RefitEvery = 100)

}
names(luGASRoll) <- colnames(dji30ret)
\end{CodeInput}
\end{CodeChunk}
and:
\begin{CodeChunk}
\begin{CodeInput}
mGASRoll <- MultiGASRoll(data = dji30ret[, c("CAT", "MMM", "PFE")],
                        GASSpec = mGASSpec,
                        ForecastLength = 3000,
                        RefitEvery = 100)
\end{CodeInput}
\end{CodeChunk}
for the univariate and multivariate cases, respectively.

Let us now compare the ability of GAS and GARCH models in predicting the one--step ahead distribution using so--called \emph{scoring rules}, which compare the predicted density with the ex--post realized value of the return and deliver a score which defines a ranking across the alternative models at each point in time \citep{gneiting_etal.2007}.
Generally, we define $S_{t+1}\bydef S(y_{t+1}, p(y_{t+1};\widehat\btheta_{t+1}))$ as the score at time $t+1$ for having predicted $p(y_{t+1};\widehat\btheta_{t+1})$ when $y_{t+1}$ has been realized. We consider two widely used scoring rules:
\begin{itemize}
  \item The average weighted Continuous Ranked Probability Score (wCRPS):
  \begin{equation}\label{eq:crsp}
    \overline{wCRPS} \bydef \frac{1}{H}\sum_{t = T}^{T+H-1}\int_{-\infty}^{\infty}w\left(z\right)\left(F\left(z;\widehat\btheta_{t+1}\right) - \mathbb{I}_{\{y_{t+1}<z\}}\right)^2\mathrm{d}z,
  \end{equation}

  where $w\left(z\right)$ is a weight function that emphasizes regions of interest of the predictive distribution, such as the tails or the center. Similarly to \citet{gneiting_ranjan.2012}, we consider the cases of: (i) a weighting that gives equal emphasis to all the parts of the distribution; $w\left(z\right) = 1$, (ii) a weighting that focuses on the center; $w\left(z\right) = \phi_{a,b}\left(z\right)$; (iii) a weighting that focuses on the tails; $w\left(z\right) = 1 - \phi_{a,b}\left(z\right)/\phi_{a,b}\left(0\right)$, (iv) a weighting that focuses on the right tail; $w\left(z\right) = \Phi_{a,b}\left(z\right)$, and (v) a weighting that focuses on the left tail $w\left(z\right) = 1 - \Phi_{a,b}\left(z\right)$. The  functions $\phi_{a,b}\left(z\right)$ and $\Phi_{a,b}\left(z\right)$ are the \textit{pdf} and \textit{cdf} of a Gaussian distribution with mean $a$ and standard deviation $b$, respectively. The label \code{uniform} represents the case where equal emphasis is given to all the parts of the distribution.
   \item The average Negative Log Score (NLS):

  \begin{equation}\label{eq:logscore}
    \overline{NLS} \bydef -\frac{1}{H}\sum_{t = T}^{T+H-1} \log p(y_{t+1};\widehat\btheta_{t+1}).
  \end{equation}
  Consistent with \citet{gneiting_etal.2007}, we specify the Negative Log Score such that the \qmo direction\qmcsp between the two scoring rules is the same, \emph{i.e.}, forecasts with lower $\overline{NLS}$ and lower $\overline{wCRPS}$ are preferred.
\end{itemize}
The two aforementioned scoring rules can be easily evaluated using the \code{BacktestDensity()} function available in the \proglang{R} package \pkg{GAS}. The \code{BacktestDensity()} function accepts an object of the class \code{uGASRoll}, and returns a list with two elements: (i) the averages negative LS and wCRPS as in~\eqref{eq:logscore} and~\eqref{eq:crsp}, and (ii) their values at each point in time. Additional arguments are:
\begin{itemize}
    \item \code{lower}: \code{numeric} representing the lower bound used to approximate~\eqref{eq:crsp} by Monte Carlo integration as detailed in \cite{gneiting_ranjan.2012}.
    \item \code{upper}: \code{numeric} as \code{lower} but for the upper bound.\footnote{The two arguments \code{lower} and \code{upper} coincide with $y_l$ and $y_u$ in Equation~16 of \cite{gneiting_ranjan.2012}, respectively. These are two numeric objects with no default value, \emph{i.e.}, the user have to define these values according to her research design.}
    \item \code{K}: \code{numeric} integer representing the number of points used to discretize the integral in~\eqref{eq:crsp}.\footnote{Equals to $I$ in Equation~16 of \cite{gneiting_ranjan.2012}.} By default \code{K = 1000},
\end{itemize}
plus the two \code{numeric} arguments, \code{a} and \code{b}, representing $a$ and $b$ in the weight functions, by default \code{a = 0.0} and \code{b = 1.0}.\footnote{These values can be chosen in order to target some \qmo optimal\qmcsp prediction level, or to add more flexibility and focus on specific parts the predictive distribution; see \cite{gneiting_ranjan.2012}.}

\begin{table}[!t]
	\centering
  \setlength\tabcolsep{15pt}
	\begin{tabular}{lrrrrrr}
		\toprule
Asset & NLS & Uniform & Center & Tails & Tails--r & Tails--l\\
\cmidrule(lr){1-1}\cmidrule(lr){2-7}
AA & $-1.99^b$ & $-2.39^a$ & $-2.39^a$ & $-1.45^c$ & $-2.46^a$ & $-2.32^b$ \\
AXP & $-2.85^a$ & $-2.25^b$ & $-2.25^b$ & $0.20^{~}$ & $-2.25^b$ & $-2.25^b$ \\
BA & $-1.54^c$ & $-2.00^b$ & $-2.00^b$ & $-0.40^{~}$ & $-2.00^b$ & $-1.99^b$ \\
BAC & $-3.16^a$ & $-1.49^c$ & $-1.50^c$ & $0.64^{~}$ & $-1.25^{~}$ & $-1.73^b$ \\
C & $-4.07^a$ & $-3.14^a$ & $-3.14^a$ & $-0.29^{~}$ & $-3.18^a$ & $-3.09^a$ \\
CAT & $-5.47^a$ & $-5.38^a$ & $-5.38^a$ & $-1.94^b$ & $-5.32^a$ & $-5.43^a$ \\
CVX & $0.71^{~}$ & $1.13^{~}$ & $1.13^{~}$ & $1.27^{~}$ & $1.12^{~}$ & $1.13^{~}$ \\
DD & $-1.34^c$ & $-0.66^{~}$ & $-0.66^{~}$ & $-0.27^{~}$ & $-0.62^{~}$ & $-0.69^{~}$ \\
DIS & $-2.17^b$ & $-1.76^b$ & $-1.76^b$ & $-0.40^{~}$ & $-1.78^b$ & $-1.74^b$ \\
GE & $-3.81^a$ & $-4.22^a$ & $-4.22^a$ & $-1.73^b$ & $-4.23^a$ & $-4.20^a$ \\
GM & $0.09^{~}$ & $-0.44^{~}$ & $-0.45^{~}$ & $0.45^{~}$ & $-0.51^{~}$ & $-0.37^{~}$ \\
HD & $-3.67^a$ & $-2.89^a$ & $-2.89^a$ & $-3.32^a$ & $-2.89^a$ & $-2.88^a$ \\
HPQ & $-4.38^a$ & $-4.44^a$ & $-4.44^a$ & $-3.42^a$ & $-4.42^a$ & $-4.45^a$ \\
IBM & $-3.92^a$ & $-4.03^a$ & $-4.03^a$ & $-2.95^a$ & $-4.00^a$ & $-4.06^a$ \\
INTC & $-3.16^a$ & $-1.72^b$ & $-1.72^b$ & $-1.48^c$ & $-1.80^b$ & $-1.65^b$ \\
JNJ & $-3.81^a$ & $-2.08^b$ & $-2.08^b$ & $-0.41^{~}$ & $-2.01^b$ & $-2.16^b$ \\
JPM & $-1.97^b$ & $-1.93^b$ & $-1.93^b$ & $-0.23^{~}$ & $-1.87^b$ & $-1.99^b$ \\
AIG & $-0.22^{~}$ & $0.94^{~}$ & $0.95^{~}$ & $0.68^{~}$ & $0.76^{~}$ & $1.05^{~}$ \\
KO & $-3.36^a$ & $-3.03^a$ & $-3.03^a$ & $-0.95^{~}$ & $-3.05^a$ & $-3.00^a$ \\
MCD & $-2.04^b$ & $-1.96^b$ & $-1.96^b$ & $-1.02^{~}$ & $-1.96^b$ & $-1.95^b$ \\
MMM & $-4.22^a$ & $-4.62^a$ & $-4.62^a$ & $-2.26^b$ & $-4.62^a$ & $-4.62^a$ \\
MRK & $-3.82^a$ & $-5.10^a$ & $-5.10^a$ & $-3.70^a$ & $-5.18^a$ & $-5.02^a$ \\
MSFT & $-3.24^a$ & $-2.76^a$ & $-2.76^a$ & $-2.25^b$ & $-2.79^a$ & $-2.73^a$ \\
PFE & $-4.86^a$ & $-4.98^a$ & $-4.98^a$ & $-3.21^a$ & $-5.02^a$ & $-4.94^a$ \\
PG & $-1.97^b$ & $-1.99^b$ & $-1.99^b$ & $-2.07^b$ & $-2.01^b$ & $-1.97^b$ \\
T & $-0.31^{~}$ & $-0.16^{~}$ & $-0.16^{~}$ & $0.46^{~}$ & $-0.14^{~}$ & $-0.18^{~}$ \\
UTX & $-1.50^c$ & $-1.52^c$ & $-1.52^c$ & $-0.80^{~}$ & $-1.57^c$ & $-1.46^c$ \\
VZ & $-2.22^b$ & $-2.13^b$ & $-2.13^b$ & $-2.41^a$ & $-2.13^b$ & $-2.14^b$ \\
WMT & $-2.15^b$ & $-1.83^b$ & $-1.83^b$ & $-0.13^{~}$ & $-1.88^b$ & $-1.78^b$ \\
XOM & $0.17^{~}$ & $0.39^{~}$ & $0.39^{~}$ & $1.26^{~}$ & $0.42^{~}$ & $0.37^{~}$ \\
		\bottomrule
	\end{tabular}
	\caption{Test statistics for the \cite{diebold_mariano.1995} test of equal performance between the series of negative Log Scores and weighed Continuous Ranked Probability Scores for univariate GAS and GARCH models across the out--of--sample logarithmic returns of Dow Jones 30 constituents. Negative values indicate that GAS models report more accurate predictions of the one--step ahead conditional distribution while positive values favour GARCH. The apexes $a,b$ and $c$ represent  rejection of the null hypothesis of Equal Predictive Ability at the $1\%, 5\%$ and $10\%$ confidence levels, respectively. The out--of--sample period spans from January 27th, 1991, to February 3rd, 2009 for a total of 3,000 observations.}
	\label{tab:BackDist}
\end{table}

In our case, in order to evaluate $\overline{NLS}$ and $\overline{wCRPS}$ for the first asset we can simply run:
\begin{CodeChunk}
\begin{CodeInput}
R> DensityBacktest <- BacktestDensity(luGASRoll[[1]],
                                      lower = -1.0, upper = 1.0)
R> DensityBacktest$average
\end{CodeInput}
\begin{CodeOutput}
        LS    uniform     center      tails     tail_r     tail_l
-2.389e+00  1.329e-02  5.300e-03  7.310e-06  6.645e-03  6.648e-03
\end{CodeOutput}
\end{CodeChunk}
where \code{lower = -1.0} and \code{upper = 1.0}.\footnote{Chosen \code{lower} and \code{upper} values define a proper range for log--returns not in percentage points as the one considered here.}

Table~\ref{tab:BackDist} reports the test statistics for the \citet{diebold_mariano.1995} (DM) test of equal performance between the series of Negative Log Scores and weighed Continuous Ranked Probability Scores for univariate GAS and GARCH models across the out--of--sample period. Negative values indicate that GAS models generate more accurate predictions of the one--step ahead conditional distribution while positive values favour GARCH. We found that, for almost all the series, GAS outperforms GARCH at very high confidence levels according to both NLS and wCRPS. Interestingly, our results suggest that GAS delivers more accurate results whatever part of the conditional distribution the $\overline{wCRPS}$ emphasizes.

\begin{figure}[!t]
\centering
\includegraphics[width=1\textwidth]{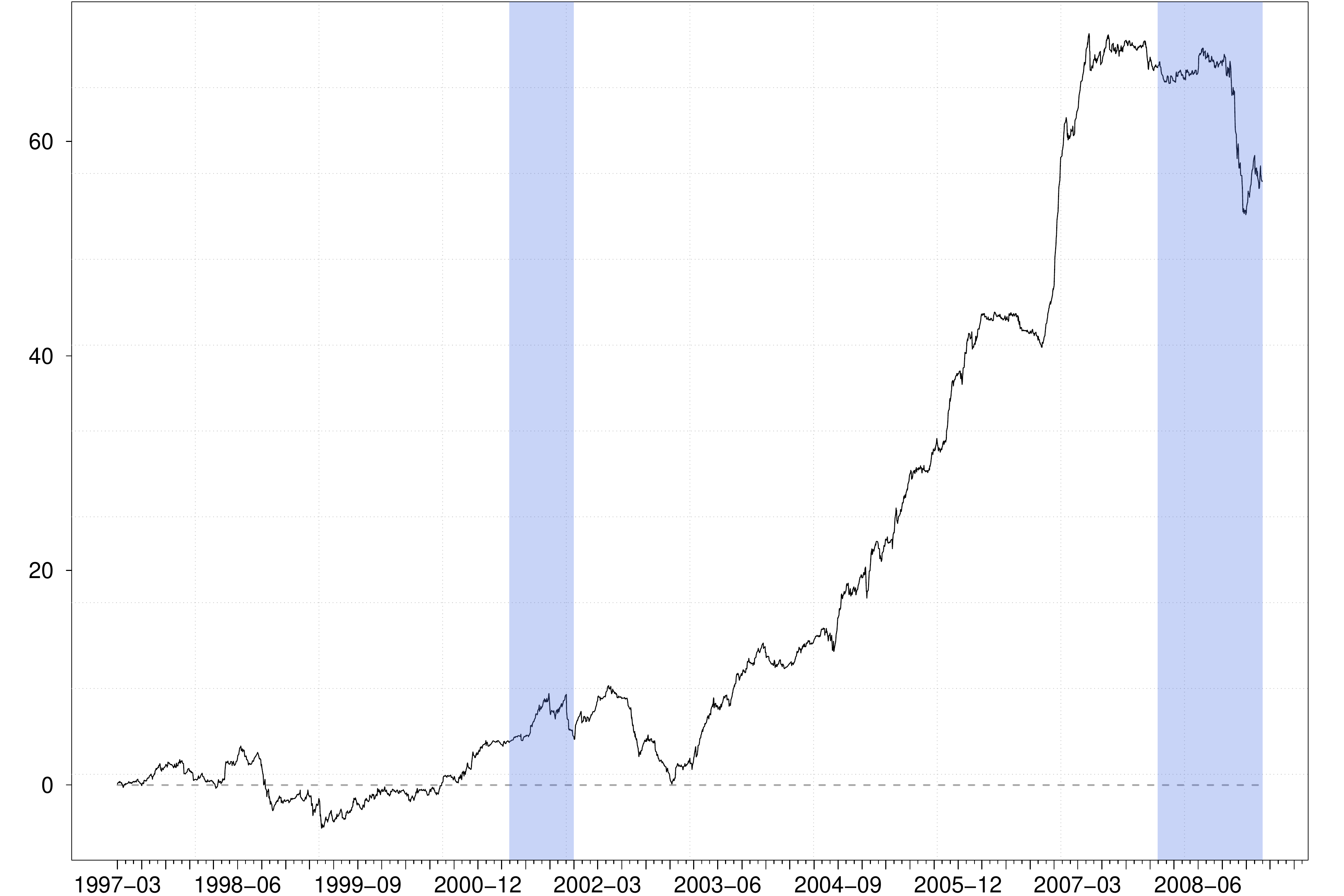}
\caption{Cumulative out--of--sample Log Score differences between the multivariate \Student GAS and the DCC(1,1) model of \citet{engle.2002} with multivariate \Student errors. Periods when the plot line slopes upward represent periods in which GAS outperforms GARCH, while downward--sloping segments indicate periods when the GARCH forecast is more accurate.  The blue shaded area represents periods of recession in the US economy according to the \qmo USREC\qmcsp series available from the Federal Reserve Bank of St. Louis web site at  \url{https://fred.stlouisfed.org/series/USREC}.}
\label{fig:cumlogscore}
\end{figure}

For the multivariate analysis we only consider $\overline{NLS}$. In this case, the DM test statistic is $-4.15$, which strongly favours the GAS model against the DCC specification. To further investigate this result, we report in Figure~\ref{fig:cumlogscore} the Cumulative sum of the differences between the Log Scores (CLS) of GAS and DCC defined as:
\begin{equation*}
  CLS_{T:T+l}^{GAS\vert DCC} \bydef \sum_{t = T}^{t = T+l-1}\log p\left(\by_{t+1};\widehat{\btheta}_{t+1}^{GAS}\right) - \log p\left(\by_{t+1};\widehat{\btheta}_{t+1}^{DCC}\right)\,,
\end{equation*}
where $p\left(\by_{t+1};\widehat{\btheta}_{t+1}^{GAS}\right)$ and $p\left(\by_{t+1};\widehat{\btheta}_{t+1}^{DCC}\right)$ are the densities predicted from GAS and DCC evaluated in $\by_{t+1}$, respectively. The series of Log Scores for the multivariate GAS models is available in the output of the \code{BacktestDensity()} function, or can be extracted using the \code{LogScore} method defined for \code{mGASRoll} objects:
\begin{CodeChunk}
\begin{CodeInput}
R> LS_MGAS <- LogScore(mGASRoll)
\end{CodeInput}
\end{CodeChunk}

Looking at Figure~\ref{fig:cumlogscore}, periods when the plot line slopes upward represent periods in which GAS outperforms DCC, while downward--sloping segments indicate periods when the DCC forecast is more accurate. From this plot, we clearly understand the output of the DM test. Interestingly, we found that GAS starts dominating DCC after 2000.

\section[Conclusion]{Conclusion}\label{sec:conclusion}

This article introduced the \proglang{R} package \pkg{GAS} for simulating, estimating and forecasting time--varying parameter models under the Generalized Autoregressive Score framework. It allows   practitioners in many scientific areas to perform their applied research using GAS models in a  user--friendly environment.


We introduced the model specification in a general way and illustrated the package usage. In particular, we performed an empirical application using financial data in which we compared the performance of univariate and multivariate GAS and GARCH models. Given the flexibility of GAS models and the availability of several statistical distributions in the \pkg{GAS} package, a number of different applications can be easily handled, such as: (i) the analysis of integer valued time series using the Poisson GAS model (\code{poi}), (ii) the analysis of (0,1)--bounded time series using the Beta GAS model (\code{beta}), (iii) the analysis of strictly positive time series with an inverse location/scale dependence using the Gamma GAS model (\code{gamma}).


Finally, if you use \proglang{R} or \pkg{GAS}, please cite the software in publications.

\section*{Computational details}

The results in this paper were obtained using \proglang{R} 3.2.3 \citep{R} with the
packages: \pkg{GAS} version 0.1.4 \citep{GAS}, \pkg{MASS} version 7.3-45 and \citep{MASS.2002,MASS},
\pkg{Rcpp} version 0.12.7 \citep{Rcpp.2011,Rcpp}, \pkg{RcppArmadillo} version 0.7.400.2.0 \citep{RcppArmadillo.2014,RcppArmadillo}, \pkg{Rsolnp} version 1.16 \citep{Rsolnp}, \pkg{xts} version 0.9-7 \citep{xts} and \pkg{quantmod} version 0.4-6 \citep{quantmod}.
\proglang{R} itself and all packages used are available
from \proglang{CRAN} at \url{http://CRAN.R-project.org/}. The
package \pkg{GAS} is available from the CRAN repository at \url{https://cran.r-project.org/package=GAS}. The version under development is available in \proglang{GitHub} at \url{https://github.com/LeopoldoCatania/GAS}.
Computations were performed on
a Genuine Intel\textregistered{} quad core CPU i7--3630QM 2.40Ghz processor. Code outputs were
obtained using \code{options(digits = 4, max.print = 40, prompt = "R> ")}

The folder \code{inst/doc} inside the \pkg{GAS} package tarball contains additional technical documentations. A step by step guide on how to add a new statistical distribution in the \pkg{GAS} package is reported in the file \code{AddNewDistribution.pdf}.

\section*{Acknowledgments}

The authors acknowledge Google for financial support via the
Google Summer of Code 2016  project
"GAS"; see \url{https://summerofcode.withgoogle.com/projects/#4717600793690112}.
Any remaining errors or shortcomings are the authors' responsibility.


%
\clearpage
\appendix
%
\section[The GAS model with conditional Student-t distribution]{The GAS model with conditional \Student distribution}\label{sec:gas_t}

Let us consider the case where the distribution of the univariate random variable $y_t\in\Re$, conditionally on $\by_{1:t-1}$, is \Student with location $\mu_t$, scale $\phi_t$ and $\nu_t$ degrees of freedom, \emph{i.e.}, $\btheta_t = (\mu_t,\phi_t,\nu_t)^\prime$ and:
\begin{equation}~\label{eq:std}
  p(y_t;\btheta_t) \bydef \frac{\Gamma\left(\frac{\nu_t + 1}{2}\right)}{\Gamma\left(\frac{\nu_t}{2}\right)\phi_t\sqrt{\pi\nu_t}}\left(1 + \frac{\left(y_t - \mu_t\right)^2}{\nu_t\phi_t^2}\right)^{-\frac{\nu_t + 1}{2}} \,.
\end{equation}
As will become clear, the score corresponding to the \Student distribution has the advantage of dampening the effect of extreme observations on the future volatility, when
the \Student has sufficiently fat tails. It has been used by \citet{creal_etal.2013} and \citet{lucas_zhang.2016} under the name tGAS, and by \citet{harvey.2013} and \citet{harvey_luati.2014} under the name Beta--t--EGARCH.

Differentiating the logarithm of~\eqref{eq:std} with respect to $\btheta_t$ leads to the score vector $\bnabla_t(y_t,\btheta_t)=(\nabla_t^\mu,\nabla_t^\phi,\nabla_t^\nu)'$, with:
\begin{align*}
\begin{split}
  \nabla_t^\mu &\bydef \frac{(\nu_t + 1)(y_t - \mu_t)}{\nu_t\phi_t\left(1 + \frac{\left(y_t - \mu_t\right)^2}{\nu_t\phi_t}\right)} \\
  \nabla_t^\phi &\bydef \frac{\left(\nu_t + 1\right)\left(y_t - \mu_t\right)^2}{2\nu_t\phi_t^2\left(1 + \frac{\left(y_t - \mu_t\right)^2}{\nu_t\phi_t}\right)} - \frac{1}{\phi_t}\\
  \nabla_t^\nu &\bydef \frac{1}{2}\psi\left(\frac{\nu_t + 1}{2}\right) - \frac{1}{2}\psi\left(\frac{\nu_t}{2}\right) - \frac{1}{2\nu_t}\\
  &- \frac{1}{2}\log\left(1 + \frac{\left(y_t - \mu_t\right)^2}{\nu_t\phi_t}\right) + \frac{\left(\nu_t + 1\right)\left(y_t - \mu_t\right)^2}{2\nu_t^2\phi_t\left(1 + \frac{\left(y_t - \mu_t\right)^2}{\nu_t\phi_t}\right)} \,,
\end{split}
\end{align*}
where $\psi(\cdot)$ is the Digamma function. Without loss of generality, let us consider the case where $\gamma = 0$ with no reparametrization, \emph{i.e.}, $\btheta_t = \widetilde\btheta_t$. The results when $\gamma \neq 0$ and a mapping function $\Lambda(\cdot)$ for $\btheta_t$ is introduced are qualitatively the same. Clearly, what controls for the response to extreme observations in the conditional score $\bnabla_t(y_t,\btheta_t)$ is the degree of freedom parameter $\nu_t$. When $\nu_t$ is small, say $\nu_t = 3$, the conditional distribution of $y_t$ has high probability mass in the tails, which means that extreme observations, which would be considered outliers under the conditionally normal distribution, are likely to be observed.

If we introduce the following mapping function for the unrestricted vector of parameter $\widetilde\btheta_t = (\widetilde\mu_t,\widetilde\phi_t,\widetilde\nu_t)'$:
\begin{equation*}
  \Lambda(\widetilde\btheta_t) \bydef \begin{cases}
                                          \mu_t  \bydef& \!\!\widetilde\mu_t \\
                                          \phi_t \bydef& \!\!\exp(\widetilde\phi_t) \\
                                          \nu_t \bydef& \!\!\exp(\widetilde\nu_t) + c\,,
                                        \end{cases}
\end{equation*}
with $c=2$ in order to ensure the existence of $V_{t-1}\left[y_t\right]$, then the GAS updating step for $\btheta_t$ when $\gamma = 0$ takes the form:
\begin{align}~\label{eq:update_studentt}
\begin{split}
  \btheta_{t+1} &\bydef  \Lambda(\widetilde\btheta_{t+1})\\
  \widetilde\btheta_{t+1} &\bydef \bkappa + \bA\mathcal{J}(\widetilde\btheta_t)'\bnabla_t(y_t,\btheta_t) + \bB\widetilde\btheta_t \,,
\end{split}
\end{align}
where $\bkappa\bydef(\kappa_\mu, \kappa_\phi, \kappa_\nu)'$, $\bA \bydef \diag(a_\mu, a_\phi, a_\nu)$ and $\bB \bydef \diag(b_\mu, b_\phi, b_\nu)$. In this particular case, the Jacobian matrix $\mathcal{J}(\widetilde\btheta_t)$ takes the form:
\begin{equation*}
  \mathcal{J}(\widetilde\btheta_t) = \begin{pmatrix}
                                             1 & 0 & 0 \\
                                             0 & \exp(\widetilde\phi_t) & 0 \\
                                             0 & 0 & \exp(\widetilde\nu_t)
                                           \end{pmatrix}\,.
\end{equation*}

Constraints on the evolution of the GAS parameters can be easily considered by fixing the values of the $\bA$ and $\bB$ elements. For example, if the constraint $\nu_t=\nu$ has to be imposed, we set $a_\nu =  b_\nu = 0$ during the (log-)likelihood maximization.

\section[Mapping functions]{Mapping functions}\label{sec:mapping}

Now we briefly discuss the choice of the mapping function $\Lambda(\cdot)$ for GAS models. We indicate the $i$--th element of $\btheta_t$ and $\widetilde\btheta_t$ as $\theta_{i,t}$ and $\widetilde\theta_{i,t}$, respectively. Analogously, we refer to the $i$--th element of the vector--valued mapping function $\Lambda(\cdot)$ as $\lambda_{i}(\cdot)$, such that $\lambda_i(\widetilde\theta_{i,t})=\theta_{i,t}$.

Generally, there are three types of constraints we want to impose on $\theta_{i,t}$:
\begin{itemize}
  \item[1)] $\theta_{i,t}>c,\quad c\in\Re$
  \item[2)] $\theta_{i,t}\in\left(a,b\right)$, for $a,b\in\Re$ and $b>a$
  \item[3)] $\theta_{i,t}\in\left(a,b\right)\vert \btheta_t \in\Theta$ for $a,b\in\Re$ and $b>a$~,
\end{itemize}
the additional case when $\theta_{i,t}\in\Re$, and thus $\widetilde\theta_{i,t} = \theta_{i,t}$, implicitly requires that $\lambda_i:\Re\to\Re$ is the identity function.

The first case, $\theta_{i,t}>c$,\footnote{The case $\theta_{i,t}<c$ follows immediately.} covers the situation where, for example, $\theta_{i,t}$ is a scale parameter and, consequently, its positiveness has to be imposed (\emph{i.e.}, $c=0$). In this case, $\lambda_i:\Re\to\left[c,\infty\right)$, and the exponential link function, defined as:
\begin{equation}\label{eq:exponential_link}
  \theta_{i,t} = \exp(\widetilde\theta_{i,t}) + c\,,
\end{equation}
can be employed. The second case, $\theta_{i,t}\in\left(a,b\right)$, covers the situation where, for example, $p\left(\cdot;\btheta_t\right)$, is the asymmetric \Student distribution of \cite{zhu_galbraith.2010}, and $\theta_{i,t}$ is its skew parameter defined in $\left(0,1\right)$. In the more general case we have $\lambda_i:\Re\to\left(a, b\right)$, and thus, the modified logistic function:
\begin{equation}\label{eq:logistic_link}
  \theta_{i,t} = a + \frac{b - a}{1 + \exp(-\widetilde\theta_t)} \,,
\end{equation}
can be employed. The last case, $\theta_{i,t}\in\left(a,b\right)\vert \btheta_t \in\Theta$, is more complicated and covers the situation where, for example, $p\left(\cdot;\btheta_t\right)$ is a multivariate Gaussian distribution and $\theta_{i,t}$ is one element of its correlation matrix $\bR_{t}$. Clearly, in this case $\theta_{i,t}\subseteq\left[-1,1\right]$, with the equivalence corresponding to the case $N=2$. For the more general case $N>1$, we need to ensure that $\bR_t$ is positive definite, \emph{i.e.}, $\bx^\prime\bR_t\bx>0,\forall\bx\in\Re^N$. Following \cite{creal_etal.2011}, we employ the hyperspherical coordinates transformation originally proposed by \cite{pinheiro_bates.1996} and subsequently discussed in \cite{jaeckel_rebonato.1999}, \cite{rapisarda_etal.2007} and \cite{pourahmadi_wang.2015}. We define the general $(h,k)$--th lower diagonal element of $\bR_t$ as $\rho_{hk,t}=\theta_{i,t}$ for $h>k$, $h<N$ and $\widetilde\rho_{hk,t}=\widetilde\theta_{i,t}$, for $i=1,\dots,N\left(N-1\right)/2$. \cite{pourahmadi_wang.2015} show that:
\begin{equation*}
  \rho_{hk,t} = c_{h1,t}c_{k1,t} + \sum_{m=2}^{h-1}c_{hm,t}c_{km,t}\prod_{l=1}^{m-1}s_{hl,t}s_{kl,t} + c_{hk,t}\prod_{l=1}^{h-1}s_{hl,t}s_{kl,1}\quad 1\le h<k\le N\,,
\end{equation*}
where $c_{hk,t} \bydef \cos\left(\widetilde\rho_{hk,t}\right)$ and $s_{hk,t} \bydef \sin\left(\widetilde\rho_{hk,t}\right)$ for all $1\le h<k\le N$ ensure that $\bR_t \bydef \{\rho_{ij,t}\}_{i,j=1}^N$ is a proper correlation matrix. 

These three specifications for $\lambda_i\left(\cdot\right)$ cover all the cases considered in this article and in the \proglang{R} package \pkg{GAS}. Additional information are reported in the \proglang{R} documentation. Fore details on $\Lambda\left(\cdot\right)$ and $\Lambda^{-1}\left(\cdot\right)$; see \code{help("UniMapParameters")} and \code{help("UniUnmapParameters")} in the univariate case and \code{help("MultiMapParameters")} and  \code{help("MultiUnmapParameters")} in the multivariate case.

\end{document}